\documentclass[aps,prappl,twocolumn,groupedaddress,showkeys,showpacs,superscriptaddress,floatfix]{revtex4-1}
\usepackage{epsfig}
\usepackage{multirow}
\usepackage{amsmath, amssymb,mathtools}
\usepackage{color}
\usepackage{graphicx}
\usepackage{dcolumn}
\usepackage{bm}
\usepackage{subfigure}
\usepackage[utf8]{inputenc}
\usepackage[T1]{fontenc}
\usepackage{tikz}
\newcommand*\bit[1]{\tikz[baseline=(char.base)]{\node[shape=rectangle,draw,inner sep=0.8pt] (char) {\scriptsize{\texttt{#1}}};}}

\begin{document}

\title{A Josephson thermal memory}

\author{Claudio Guarcello}
\email{claudio.guarcello@nano.cnr.it}
\affiliation{NEST, Istituto Nanoscienze-CNR and Scuola Normale Superiore, Piazza San Silvestro 12, I-56127 Pisa, Italy}
\affiliation{Radiophysics Department, Lobachevsky State University, Gagarin Avenue 23, 603950 Nizhni Novgorod, Russia}
\author{Paolo Solinas}
\affiliation{SPIN-CNR, Via Dodecaneso 33, 16146 Genova, Italy}
\author{Alessandro Braggio}
\affiliation{NEST, Istituto Nanoscienze-CNR and Scuola Normale Superiore, Piazza San Silvestro 12, I-56127 Pisa, Italy}
\author{Massimiliano Di Ventra}
\affiliation{Department of Physics, University of California, San Diego, La Jolla, California 92093, USA}
\author{Francesco Giazotto}
\affiliation{NEST, Istituto Nanoscienze-CNR and Scuola Normale Superiore, Piazza San Silvestro 12, I-56127 Pisa, Italy}

\date{\today}

\begin{abstract}
We propose a superconducting thermal memory device that exploits the thermal hysteresis in a flux-controlled, temperature-biased superconducting quantum-interference device (SQUID). This system reveals a flux-controllable temperature bistability, which can be used to define two well-distinguishable thermal logic states. We discuss a suitable writing-reading procedure for these memory states. The time of the memory writing operation is expected to be on the order of $\sim0.2\;\text{ns}$, for a Nb-based SQUID in thermal contact with a phonon bath at $4.2\;\text{K}$. We suggest a non-invasive readout scheme for the memory states based on the measurement of the effective resonance frequency of a tank circuit inductively coupled to the SQUID. The proposed device paves the way for a practical implementation of thermal logic and computation. The advantage of this proposal is that it represents also an example of harvesting thermal energy in superconducting circuits.
\end{abstract}


\maketitle



\section{Introduction}
\label{Intro}\vskip-0.2cm

Computing by employing heat rather than electricity would offer several advantages compared to standard electronic devices, not least that the unavoidable heat dissipation, intrinsically produced in any computational scheme, may be used as an advantage, rather than an hindrance. On a practical side, additional operations may be done without adding extra heat dissipation. Of course, realizing such a concept requires components that have always been attributed to electronics~\cite{Li12,Skl15,Ben16}. Indeed, in the past few years the thermal counterpart of conventional diodes, transistors, memories, and logic elements have been proposed and discussed thus offering a path to a type of unconventional computing with heat~\cite{Li04,Ote10,Li12,Ben13,Ben14,Kub14,Ord16,Ben16}. 

In order to process and store information by phononic heat currents, both thermal logic and thermal memories~\cite{Wan07,Wan08,Li12,Skl15} were initially conceived. However, this phonon-based thermal technology suffers from the limited speed of the heat carriers, i.e., the acoustic phonons, which is some orders of magnitude smaller than the speed of electrons. The first solid-state thermal memory was practically demonstrated in Ref.~\cite{Xie11}.
Subsequently, optical architectures for processing and managing information via thermal photons were also developed~\cite{Kub14,Ben14,Elz14,Dya15,Ben15,Son15,Ben16,Ord16,Ito16,Elz16}. Kubytskyi \emph{et al.}~\cite{Kub14} designed a thermal memory based on far-field radiative effects, in which the time to write the memory states was predicted to be on the order of a few seconds. 
Instead, a writing time of several orders of magnitude lower, i.e., $\sim5\;\text{ms}$, was expected in near-field thermal memories~\cite{Dya15}.
Alternatively, Elzouka \emph{et al.}~\cite{Elz14,Elz16} proposed a nano thermo-mechanical memory, by varying the coupling between thermal expansion and near-field radiative heat transfer. In this case, the estimated writing time was on the order of $\sim0.05\;\text{s}$. These operational times are several orders of magnitude too long for practical applications such as thermal logic and/or computation.

\begin{figure}[b!!]
\includegraphics[width=\columnwidth]{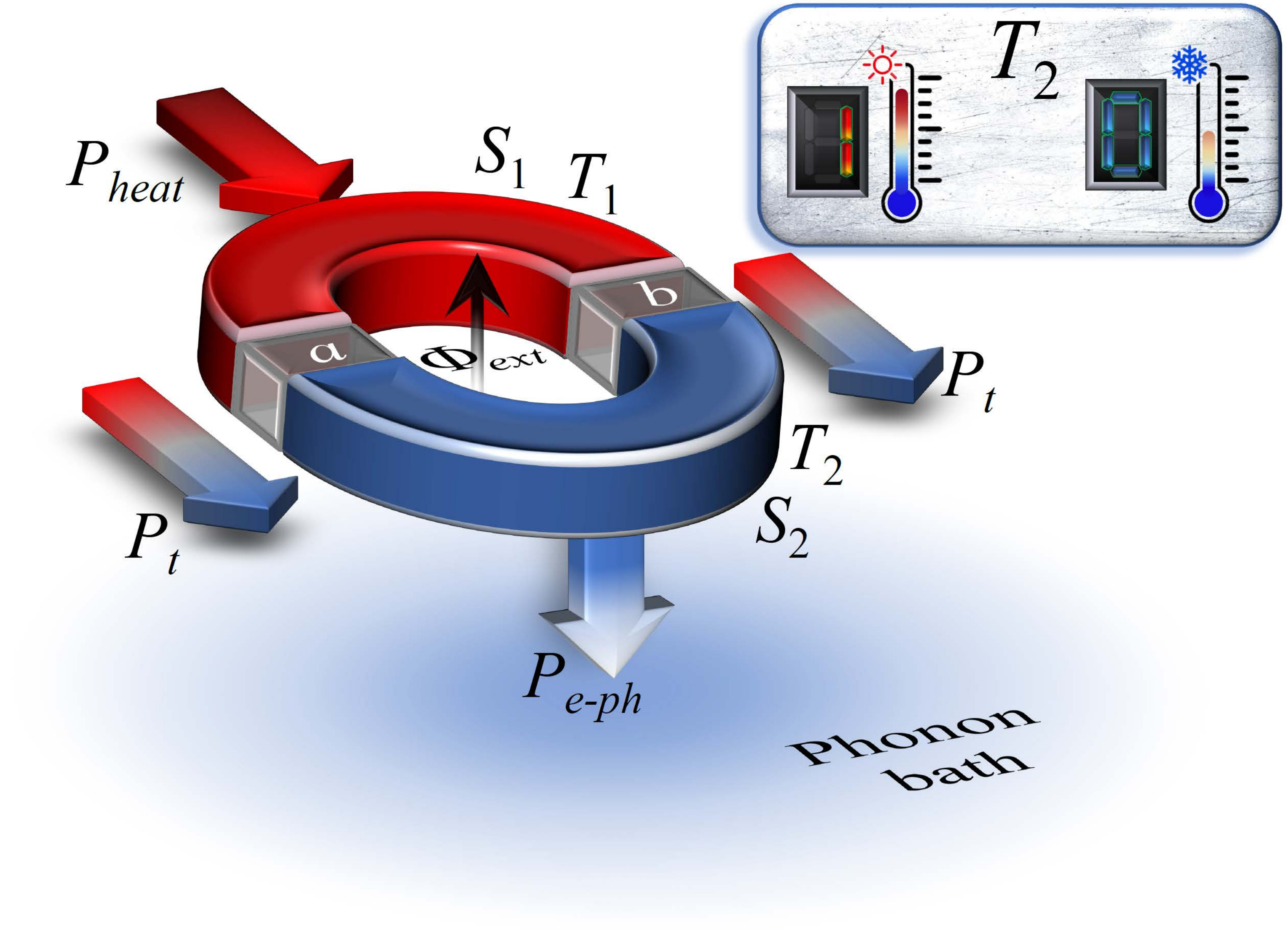}
\caption{Thermal fluxes in a magnetically driven SQUID formed by two superconductors, $S_1$ and $S_2$, at temperatures $T_1$ and $T_2$, respectively, tunnel coupled through the junctions $J_a$ and $J_b$. The
applied magnetic flux threading the SQUID loop, $\Phi_{ext}$, drives the temperature $T_2$. The heat current, $P_t$, flowing through the junctions depends on the temperatures and the total flux through the SQUID ring. $P_{e-ph}$ represents the coupling between quasiparticles in $S_2$ and the lattice phonons residing at $T_{bath}$, whereas $P_{heat}$ denotes the power injected into $S_1$ through heating probes in order to impose a fixed quasiparticle temperature $T_1$. The arrows indicate the direction of heat currents for $T_1>T_2>T_{bath}$. The temperature $T_2$ is the observable used to define the logic states 0 and 1 of the thermal memory.}
\label{Fig00}
\end{figure}

In this paper, we make a step forward in the panorama of thermal memories by introducing the concept of superconducting (Josephson) thermal memory, which uses the electronic temperature of an inductive superconducting quantum-interference device (SQUID) to define distinct thermal states. We stress that in a superconductor the electronic temperature follows dynamics appreciably faster than the phononics ones~\cite{Gia06}. Recently, the phase-coherent thermal transport~\cite{Gia12,MarGia13,Mar14,ForGia17,For17}, the negative differential thermal conductance~\cite{ForTim16}, and the hysteretical thermal behavior~\cite{Gua17} in temperature-biased Josephson devices were investigated. Here, we discuss the dynamics of a flux-controlled SQUID, delving into the hysteresis of the steady temperatures~\cite{Gua17} to define the logic states 0 and 1 of a memory device. The hysteretic behavior of the SQUID is a straight consequence of the inductive nature of the device~\cite{Cla04}, while the thermal bistability~\cite{Gua17} results from the coherent thermal transport through a temperature-biased SQUID~\cite{Mak65,Gol13}.

The paper is organized as follows. In Sec.~\ref{SQUIDModel}, the theoretical background used to describe the phase evolution of a thermally biased, magnetically driven SQUID with a non-vanishing ring inductance is discussed. In Sec.~\ref{ThermalModel}, the thermal balance equation and the heat currents are introduced. In Sec.~\ref{Results}, the thermal memory states are defined and the behaviour of the thermal memory is explored. The characteristic switching time of the memory and a fast readout scheme are also discussed. In Sec.~\ref{Conclusions}, conclusions are drawn.

\section{SQUID phase dynamics}
\label{SQUIDModel}\vskip-0.2cm

We first explore the dynamics of a magnetically driven inductive SQUID formed by two Josephson junctions (JJs), see Fig.~\ref{Fig00}. To do so, we rely on the resistively and capacitively shunted junction (RCSJ) model, describing the phase evolution of the JJs. According to the RCSJ model, the current flowing through the $i$-th JJ is given by~\cite{Cla04,Bar82,Gra16}
\begin{equation}
I_i=\frac{\Phi_0}{2\pi}C_i\ddot{\varphi_{i}}+\frac{\Phi_0}{2\pi}\frac{1}{R_i}\dot{\varphi_{i}}+I_{c_i}\sin\varphi_i,
\label{RCSJ}
\end{equation}
where $\Phi _0=h/(2e)\simeq 2\times 10^{-15}\;\textup{Wb}$ is the flux quantum ($e$ and $h$ being the electron charge and the Planck constant, respectively), and $C_i$, $R_i$, $I_{c_i}$, and $\varphi_i$ are the capacitance, the normal resistance, the critical current, and the superconducting phase difference across the $i$-th junction, respectively.

The flux quantization in a superconducting ring interspersed with two JJs imposes the constraint 
\begin{equation}
\frac{\varphi_1-\varphi_2}{2}=\pi \frac{\Phi}{\Phi_0}+\pi k,
\label{Fluxquantization}
\end{equation}
where $k$ is an integer representing the amount of enclosed flux quanta in the ring. Here, 
\begin{equation}
\Phi =\Phi_{ext}-LI_{circ}
\label{TotalFlux}
\end{equation}
 is the total magnetic flux threading the SQUID loop, where $I_{circ}$ is the circulating current and the superconducting ring inductance $L$ has a geometric contribution as well as a kinetic contribution~\cite{Cla04,Ann10}. The system is driven by the externally applied magnetic flux through the ring, $\Phi_{ext}$.

The dynamics of a SQUID formed by two JJs and biased by a current $I_{bias}$ is determined by the following system of equations~\cite{Gra16}
\begin{eqnarray}\label{RCSJa}
\frac{I_{bias}}{2}+I_{circ}=\frac{\hbar }{2e}C_1\ddot{\varphi_{1}}+\frac{\hbar}{2e}\frac{1}{R_1}\dot{\varphi_{1}}+I_{c_1}\sin\varphi_1\\
\frac{I_{bias}}{2}-I_{circ}=\frac{\hbar }{2e}C_2\ddot{\varphi_{2}}+\frac{\hbar}{2e}\frac{1}{R_2}\dot{\varphi_{2}}+I_{c_2}\sin\varphi_2,
\label{RCSJb}
\end{eqnarray}
with the constraints given by Eqs.~\eqref{Fluxquantization} and~\eqref{TotalFlux}. 

For the sake of generality, we suppose $C_1\neq C_2$ and $R_1\neq R_2$ and we introduce the quantities $C_\pm=C_1\pm C_2$, $R_\pm=R_1\pm R_2$, and $I_\pm=I_{c_1}\pm I_{c_2}$. The degree of asymmetry of the SQUID is defined as $r=I_{_{-}}/I_{_{+}}=-R_{_{-}}/R_{_{+}}$ (since $I_{c_i}\propto R_i^{-1}$~\cite{Bar82}). Once the values of $r$ and $R_1$ are chosen, the $S_2$ normal resistance can be estimated according to $R_2=R_1(1+r)/(1-r)$.

We recast Eqs.~\eqref{RCSJa} and~\eqref{RCSJb} in terms of the variables $\varphi=(\varphi_1+\varphi_2)/2$ and $\phi=(\varphi_1-\varphi_2)/2$. By the sum of Eqs.~\eqref{RCSJa} and~\eqref{RCSJb}, the following equation results
\begin{equation}
I_{bias}=\frac{\Phi_0}{2\pi}C_{_{+}}\ddot{\varphi}+\frac{\Phi_0}{2\pi}C_{_{-}}\ddot{\phi}+\frac{\Phi_0}{2\pi}\frac{1}{\mathcal{R}}\dot{\varphi}+\frac{\Phi_0}{2\pi}\frac{r}{\mathcal{R}}\dot{\phi}+I_{_{+}}f_r(\varphi,\phi),
\label{EqSQUIDSum}
\end{equation}
where $\mathcal{R}=R_1R_2/R_+$ and $f_r(\varphi,\phi)=\sin\varphi\cos\phi+r \cos\varphi\sin\phi$.

By proper normalization of the time, i.e., $\tau =2\pi \nu t$ with $\nu$ being the driving frequency, Eq.~\eqref{EqSQUIDSum} becomes
\begin{equation}
\mathcal{K}\left (\frac{\partial^2 \varphi }{\partial \tau^2}+\mathcal{C}\frac{\partial^2 \phi }{\partial \tau^2} \right )+\frac{\partial \varphi }{\partial \tau}+r\frac{\partial \phi }{\partial \tau}+\alpha\left [ f_r\left ( \varphi, \phi \right )-\delta \right ]=0,
\label{EqSQUIDSumNorm}
\end{equation}
where $\mathcal{C}=C_{_{-}}/C_{_{+}}$, $\mathcal{K}=2\pi\nu \mathcal{R}C_{_{+}}$, $\alpha =\frac{\mathcal{R} I_{_{+}}}{\nu\Phi_0}$, and $\delta =\frac{I_{bias}}{I_{_{+}}}$.
By subtracting Eqs.~\eqref{RCSJa} and \eqref{RCSJb}, one obtains
\begin{equation}
2I_{circ}=\frac{\Phi_0}{2\pi}C_{_{+}}\left (\mathcal{C}\ddot{\varphi}+\ddot{\phi} \right )+\frac{\Phi_0}{2\pi}\frac{1}{\mathcal{R}}\left (r\dot{\varphi}+\dot{\phi} \right )+I_{_{+}}g_r(\varphi,\phi),
\label{EqSQUIDSub}
\end{equation}
where $g_r(\varphi,\phi)=r\sin\varphi\cos\phi+\cos\varphi\sin\phi$.

From Eq.~\eqref{TotalFlux}, $\phi(t)-\phi_e(t)=-\frac{\beta}{2}\frac{I_{circ}}{I_{c_1}}$, where $\phi_e=\pi \Phi_{ext}/\Phi_0$ and $\beta=\frac{2\pi}{\Phi_0} L I_{c_1}$
is the SQUID hysteresis parameter~\cite{Cla04}. The higher the value of $\beta$, the greater the hysteretical response of the SQUID~\cite{Gua17}. By proper normalization of Eq.~\eqref{EqSQUIDSub} one obtains
\begin{eqnarray}\nonumber
&&\beta \mathcal{K}\left (\mathcal{C}\frac{\partial^2 \varphi }{\partial \tau^2}+\frac{\partial^2 \phi }{\partial \tau^2} \right )+\beta\left ( r\frac{\partial \varphi }{\partial \tau}+\frac{\partial \phi }{\partial \tau} \right )+\\
&&+\alpha\left \{ \beta g_r\left ( \varphi, \phi \right )+2(r+1)\left [ \phi(t)-\phi_e(t) \right ] \right \}=0.
\label{EqSQUIDSubNorm}
\end{eqnarray}

Eqs.~\eqref{EqSQUIDSumNorm} and~\eqref{EqSQUIDSubNorm} have to be solved numerically to study the behavior of the SQUID, when a non vanishing inductance, i.e., $\beta>0$, is taken into account. 
Then, the JJ's phases are calculated as $\varphi_1=\varphi+\phi$ and $\varphi_2=\varphi-\phi$.

\emph{The hysteretic parameter}. $-$ The hysteretic parameter $\beta$ is proportional to both the inductance of the superconducting ring, $L$, and the critical current $I_{c_1}$. $L$ is the sum of both a geometric and a kinetic contribution, $L_G$ and $L_{K_{ring}}$, respectively. Moreover, $\beta$ depends on the temperatures through the kinetic inductance and the critical current~\cite{Gia05,Tir08,Bos16}. Specifically,
\begin{eqnarray}
\beta(T_1,T_2)&=&\frac{2\pi}{\Phi_0}L(T_1,T_2)I_{c_1}(T_1,T_2)=\\
&=&\frac{2\pi}{\Phi_0}\left [ L_G+L_{K_{ring}}(T_1,T_2) \right ]I_{c_1}(T_1,T_2).\nonumber
\label{beta_T}
\end{eqnarray}

The BCS expression of the kinetic inductance of a superconducting strip at temperature $T$ is~\cite{Ann10}
\begin{equation}
L_K(T)=R_{sq}\frac{l}{w}\frac{\hbar}{\pi\Delta(T)\tanh\left [ \Delta(T)/(2k_BT) \right ]},
\label{L_Kwire}
\end{equation}
where $R_{sq}$ is the sheet resistance in the non-superconducting state, $l$ and $w$ are the length and the width of the strip, respectively, so that its normal resistance is $R_{strip}=\left ( \frac{l}{w} \right )R_{sq}$.

Therefore, the kinetic inductance of the superconducting ring of the SQUID with arms residing at temperatures $T_1$ and $T_2$ is
\begin{equation}
L_{K_{ring}}(T_1,T_2)= \sum_{j=1,2}R_{L_j}\frac{\hbar}{\pi\Delta(T_j)\tanh\left [ \frac{\Delta(T_j)}{2k_BT_j} \right ]},
\label{L_K}
\end{equation}
where $R_{L_j}=R_{sq}\left ( \frac{l_j}{w_j} \right )$ is the normal resistance of the $j$-th SQUID arm, with $l_j$ and $w_j$ being its length and width, respectively.

We assume a vanishing geometric inductance (i.e., $L(T_1,T_2)\simeq L_{K_{ring}}(T_1,T_2)$), since we can show that it does not crucially affects the overall behaviour. For $R_{L_1}=R_{L_2}=R_{1}$, if the temperatures are $T_1=6.5\;\text{K}$ and $T_2\in[4.2-4.5]\;\text{K}$, one obtains $\beta(T_1,T_2)\simeq1.91$. This is the value set in all the numerical calculations. Anyway, in the presence of an hysteretical behaviour of the SQUID the precise tuning of the parameters is not required to define a reliable thermal memory.

\begin{figure}[t!!]
\includegraphics[width=0.5\textwidth]{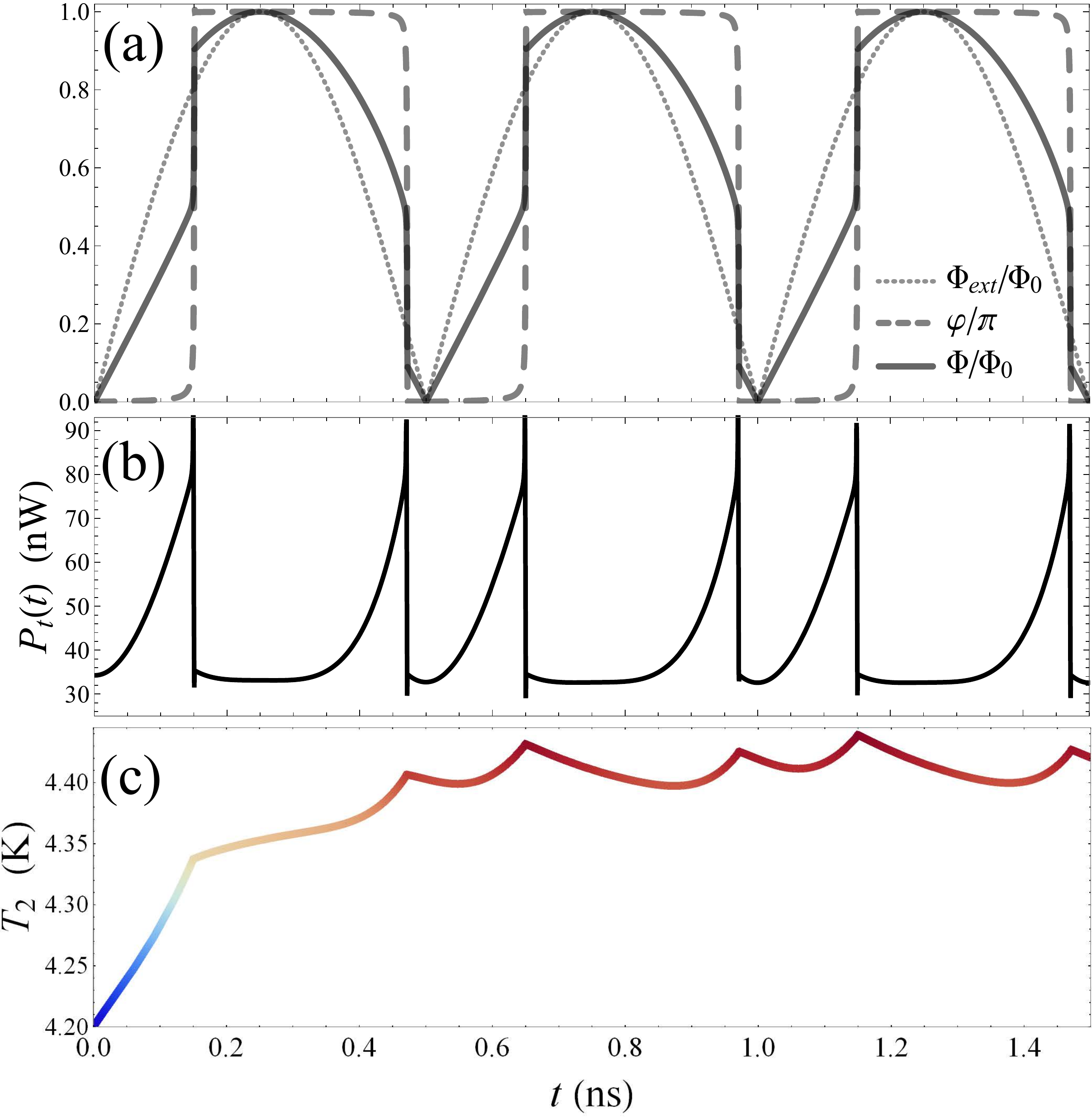}
\caption{(a) Behaviour of $\Phi/\Phi_0$ and $\varphi/\pi$ as a function of $t$, for three periods of the drive $\Phi_{ext}$. (b) and (c) Time evolution of $P(t)$ and $T_2$, respectively. The values of the other parameters are: $r=0.01$, $I_{bias}=0$, $\beta=1.91$, $R_1=10\Omega$, $C_1=C_2=10 \;\text{fF}$, $\nu=1\;\text{GHz}$, $T_1=6.5\;\text{K}$, $T_{bath}=4.2\;\text{K}$, $\mathcal{V}=10^{-1}{\mu\textup{m}}^3$, $\Sigma=3\times10^9\textup{W}\textup{m}^{-3}\textup{ K}^{-5}$, and $N_F=10^{47}\textup{ J}^{-1}\textup{ m}^{-3}$. }
\label{Fig01}
\end{figure}

\emph{The $\pi$-swap dynamics}. $-$ We investigate the behaviour of $\varphi$ and $\Phi$, shown in Fig.~\ref{Fig01}(a), of a slightly asymmetric, not-biased SQUID, i.e., $r=0.01$ and $I_{bias}=0$, driven by the magnetic flux $\Phi_{ext}=\Phi_0|\sin(2\pi\nu t)|$, with $\nu=1\;\text{GHz}$. 
The exact shape of the driving flux is not essential for the thermal bistability we will discuss.
Here, we assume $C_1=C_2=10\;\text{fF}$ and $R_1=10\;\Omega$, namely, typical values for a Nb/AlOx/Nb junction~\cite{Sol15,Pat99}, so that $R_2\simeq10.2\;\Omega$ and $I_{_{+}}\simeq0.3\;\text{mA}$. In the calculations, temperature-dependent critical currents are taken into account. 

The current circulating through the SQUID tends to compensate and screen the applied flux, see Fig.~\ref{Fig01}. In fact, according to the Faraday-Lenz law, as the external flux $\Phi_{ext}$ increases, the total flux $\Phi$ grows less rapidly than $\Phi_{ext}$, since the flux induced by the circulating current opposes $\Phi_{ext}$, see Fig.~\ref{Fig01}(a). However, as $\Phi$ approaches the critical value $\Phi_0/2$, we observe that the total flux $\Phi$ abruptly changes, i.e., a transition $k\to k\pm 1$~\cite{Gua17} takes place, and the phase difference $\varphi$ jumps from $0$ to $\pi$ (or vice versa), namely, a $\pi$-swap of $\varphi$ occurs, see Fig.~\ref{Fig01}(a). 
Then, the resulting path of alternate $\varphi$ jumps is related to the asymmetry of the device, i.e., $r>0$, which is necessary since we fix the external bias to $I_{bias}=0$~\cite{Sol15}. Indeed, in order to guarantee no effects of the electric current on the temperature difference, it is appropriate for this device to consider only non-galvanic scheme, where the SQUID is electrically isolated (floating) from external circuit.

%
%

%
\begin{figure}[t!!]
\includegraphics[width=\columnwidth]{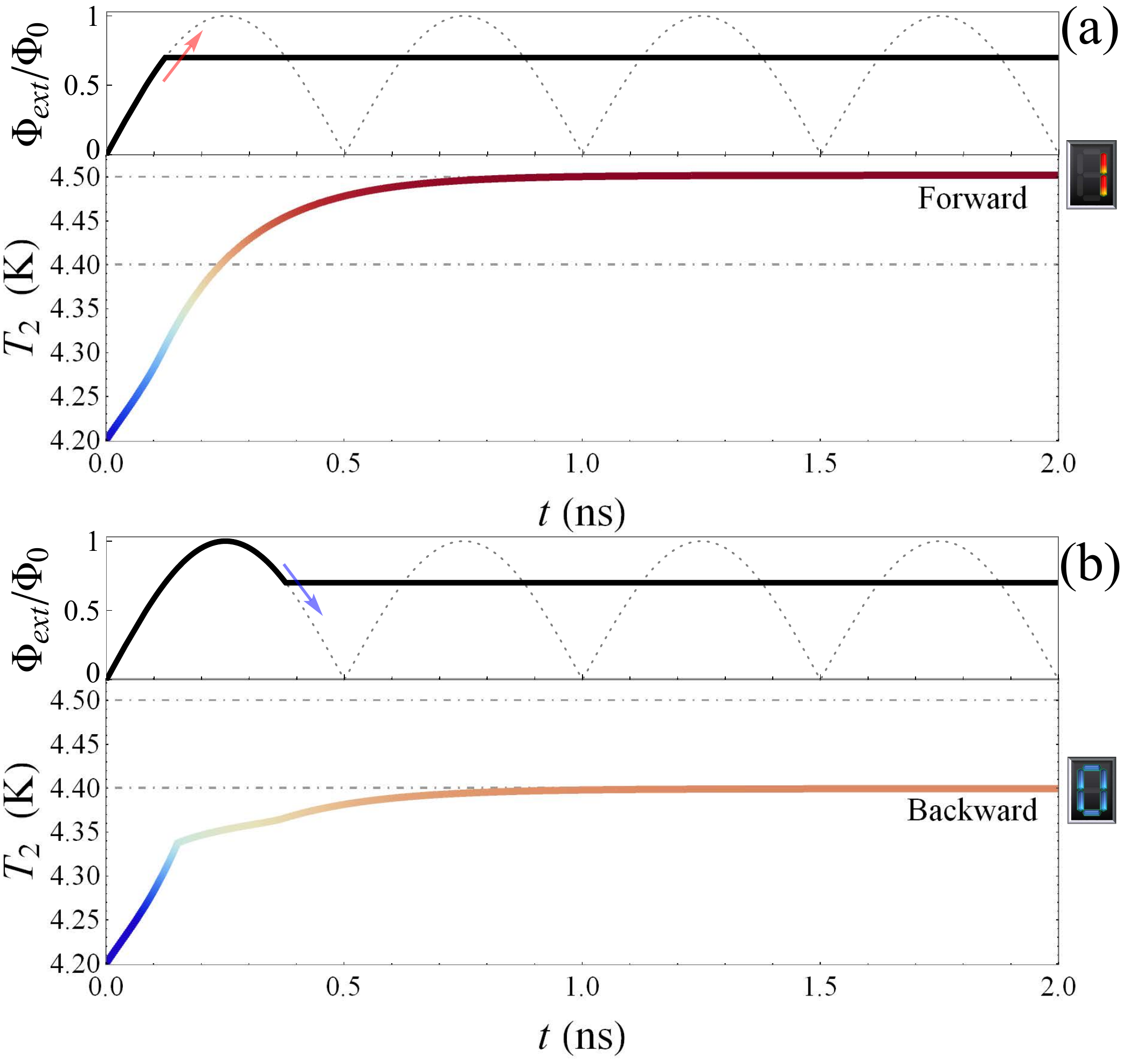}
\caption{Time evolutions of the temperature $T_2(t)$ as the external flux is kept fixed to the value $\Phi_{ref}=0.7\Phi_0$, when this value is reached within the first period during the forward (a) and backward (b) sweep of the drive, for $\nu=1\;\text{GHz}$. Dot-dashed lines represent the asymptotic values approached by $T_2$ in the two cases. In both (a) and (b), the top panel shows the driving flux $\Phi_{ext}/\Phi_0$, the dotted curve tracing the function $|\sin(2\pi\nu t)|$. A red (blue) arrow is used to indicate the increasing (decreasing) sweep direction of the drive, when its value is frozen. }
\label{Fig02}
\end{figure}

\section{Thermal dynamics}
\label{ThermalModel}\vskip-0.2cm

We now suppose a thermal bias across the SQUID, see Fig.~\ref{Fig00}. Specifically, the branch $S_1$ resides at a fixed temperature $T_1$, which is maintained by the good thermal contact with heating probes. The electronic temperature $T_2(t)$ of the branch $S_2$ is the key quantity to define our thermal memory, since it floats and can be driven by the external flux $\Phi_{ext}$. It depends on all the energy exchanges in $S_2$ which is in thermal contact with a phonon bath at a temperature $T_{bath}<T_1$. The thermal balance equation for the incoming, i.e., $P_t\left ( T_1,T_2,t \right )$, and outgoing, i.e., $P_{e-ph,2}\left ( T_2,T_{bath}\right )$, thermal powers in $S_2$ can be written as~\cite{GiaMar12}
\begin{equation}
P_t\left ( T_1,T_2,t \right )-P_{e-ph,2}\left ( T_2,T_{bath}\right )=C_v(T_2)\frac{\mathrm{d} T_2}{\mathrm{d} t}.
\label{ThermalBalanceEq}
\end{equation}
The rhs represents the variations of the internal energy of the system, with $C_v(T)=T \,\partial \mathcal{S}/\partial T$ being the heat capacity. $\mathcal{S}(T)$ is the electronic entropy of the superconductor $S_2$ and is given by~\cite{Rab08,Sol16}
\begin{equation}
\mathcal{S}(T)=-4k_BN_F\mathcal{V}\int_{-\infty}^{\infty} f(\varepsilon,T) \log f(\varepsilon,T) \mathcal{N}(\varepsilon,T) d\varepsilon.
\label{Entropy}
\end{equation}
Here, $k_B$ is the Boltzmann constant, $f ( E ,T )=1/\left (1+e^{E/k_BT} \right )$ is the Fermi distribution function, $N_F$ is the density of states at the Fermi energy, $\mathcal{V}$ is the volume of $S_2$, and $\mathcal{N}_j\left ( \varepsilon ,T_j \right )=\left | \text{Re}\left [ \frac{ \varepsilon +i\gamma_j}{\sqrt{(\varepsilon +i\gamma_j) ^2-\Delta _j\left ( T_j \right )^2}} \right ] \right |$ is the smeared BCS density of states of the $j$-th superconductor, where $\Delta_j\left ( T_j \right )$ and $\gamma_j$ are the BCS energy gap and the Dynes broadening parameter~\cite{Dyn78}, respectively.

The heat current $P_t( T_1,T_2,t)$ flowing from $S_1$ to $S_2$ reads
\begin{eqnarray}\label{Pt}\nonumber
P_t=\sum_{i=1,2}&&P_{qp,i}( T_1,T_2,V_i)+\cos\varphi_i P_{\cos,i}( T_1,T_2,V_i)+\\
&&+\sin\varphi_i P_{\sin,i}( T_1,T_2,V_i),
\end{eqnarray}
where $V_i(t)=\Phi_0/(2\pi)\dot{\varphi_i}$ is the voltage drop across the $i$-th JJ. $P_t( T_1,T_2,t)$ depends, through $\varphi_1(t)$ and $\varphi_2(t)$, on the evolution of the driving flux $\Phi_{ext}$.
In the adiabatic regime~\cite{Gol13}, the quasi-particle and the anomalous heat currents, $P_{qp,i}$, $P_{\cos,i}$, and $P_{\sin,i}$ read, respectively,~\cite{Mak65,Gol13,VirSol17}
\begin{eqnarray}\label{Pqp}\nonumber
&&P_{qp,i}(T_1,T_2,V_i )=\frac{1}{e^2R_i}\int_{-\infty}^{\infty} d\varepsilon \mathcal{N}_1 ( \varepsilon-eV_i ,T_1 )\mathcal{N}_2 ( \varepsilon ,T_2 )\\
&&\times(\varepsilon-eV_i) [ f ( \varepsilon-eV_i ,T_1 ) -f ( \varepsilon ,T_2 ) ],\\ \nonumber
&&P_{\cos,i}( T_1,T_2,V_i )=-\frac{1}{e^2R_i}\int_{-\infty}^{\infty} d\varepsilon \mathcal{N}_1 ( \varepsilon-eV_i ,T_1 )\mathcal{N}_2 ( \varepsilon ,T_2 ) \\
&&\times\frac{\Delta_1(T_1)\Delta_2(T_2)}{\varepsilon}[ f ( \varepsilon-eV_i ,T_1 ) -f ( \varepsilon ,T_2 ) ],\label{Pcos}\\\nonumber
&&P_{\sin,i}(T_1,T_2,V_i)=\frac{eV_i}{2\pi e^2R_i}\iint_{-\infty}^{\infty} d\epsilon_1d\epsilon_2 \frac{\Delta_1(T_1)\Delta_2(T_2)}{E_2}\\
&&\times\left [\frac{1-f(E_1,T_1)-f(E_2,T_2)}{\left ( E_1+E_2 \right )^2-e^2V_i^2}\text{+}\frac{f(E_1,T_1)-f(E_2,T_2)}{\left ( E_1-E_2 \right )^2-e^2V_i^2}\right ]
\end{eqnarray}
with $E_j=\sqrt{\epsilon_j^2+\Delta_j(T_j)^2}$ the Bogoliubov energies.
We observe that the anomalous terms in $P_t$ depend crucially on the junctions phases and are the terms which connect the thermal conduction of the JJs with the phase dynamics. 

In Eq.~\eqref{ThermalBalanceEq}, $P_{e-ph,2}$ represents the power loss by the quasiparticles in $S_2$ into the lattice phonons residing at $T_{bath}$~\cite{Pek09}
\begin{eqnarray}\label{Qe-ph}\nonumber
P_{e-ph,2}&=&\frac{-\Sigma \mathcal{V} }{96\zeta(5)k_B^5}\int_{-\infty }^{\infty}dEE\int_{-\infty }^{\infty}d\varepsilon \varepsilon^2\textup{sign}(\varepsilon)M_{_{E,E+\varepsilon}}\\\nonumber
&\times& \Bigg\{ \coth\left ( \frac{\varepsilon }{2k_BT_{bath}}\right ) \left [ \mathcal{F}(E,T_2)-\mathcal{F}(E+\varepsilon,T_2) \right ]\\
&-&\mathcal{F}(E,T_2)\mathcal{F}(E+\varepsilon,T_2)+1 \Bigg\},
\end{eqnarray}
where $M_{E,{E}'}=\mathcal{N}_i(E,T_2)\mathcal{N}_i({E}',T_2)\left [ 1-\Delta ^2(T_2)/(E{E}') \right ]$, $\mathcal{F}\left ( \varepsilon ,T_2 \right )=\tanh\left [ \varepsilon/(2 k_B T_2) \right ]$, $\Sigma$ is the electron-phonon coupling constant, and $\zeta$ is the Riemann zeta function. Hereafter, we impose $\mathcal{V}=10^{-1}{\mu\textup{m}}^3$, $N_F=10^{47}\textup{ J}^{-1}\textup{ m}^{-3}$, $\gamma_1=\gamma_2=10^{-4}\Delta_2(0)$, and $\Delta_1(0)=\Delta_2(0)=1.764k_BT_c$, with $T_c=9.2\;\text{K}$, namely, typical values for an Nb-based SQUID.

The temperature $T_2(t)$ is obtained by solving Eqs.~\eqref{EqSQUIDSumNorm},~\eqref{EqSQUIDSubNorm}, and~\eqref{ThermalBalanceEq} for fixed values of $T_1$ and $T_{bath}$, and represents the observable we use to encode the 0 and 1 logic states of the thermal memory.

\section{Results}
\label{Results}\vskip-0.2cm

We impose that the bath resides at $T_{bath}=4.2\;\text{K}$, and $S_1$ is at a temperature $T_1=6.5\;\text{K}$ kept fixed throughout the computation. However, the thermal logic we are going to discuss is robust to moderate fluctuations of $T_1$. The time evolution of both the power $P_t( T_1,T_2,t)$ injected into $S_2$ and $T_2(t)$ within three periods of the drive $\Phi_{ext}=\Phi_0|\sin(2\pi\nu t)|$ is shown in Figs.~\ref{Fig01}(b) and~\ref{Fig01}(c), respectively. The phase dependency of $P_t$ is clearly visible since it increases and abruptly falls in correspondence of the $\pi$-swaps of $\varphi$.
The temperature $T_2(t)$ increases from $T_2(0)=T_{bath}$ around a quasi-equilibrium value~\cite{Gua17} determined by Eq.~\eqref{ThermalBalanceEq}, with peaks in correspondence of the $P_t$'s jumps. 

%
\begin{figure}[t!!]
\includegraphics[width=\columnwidth]{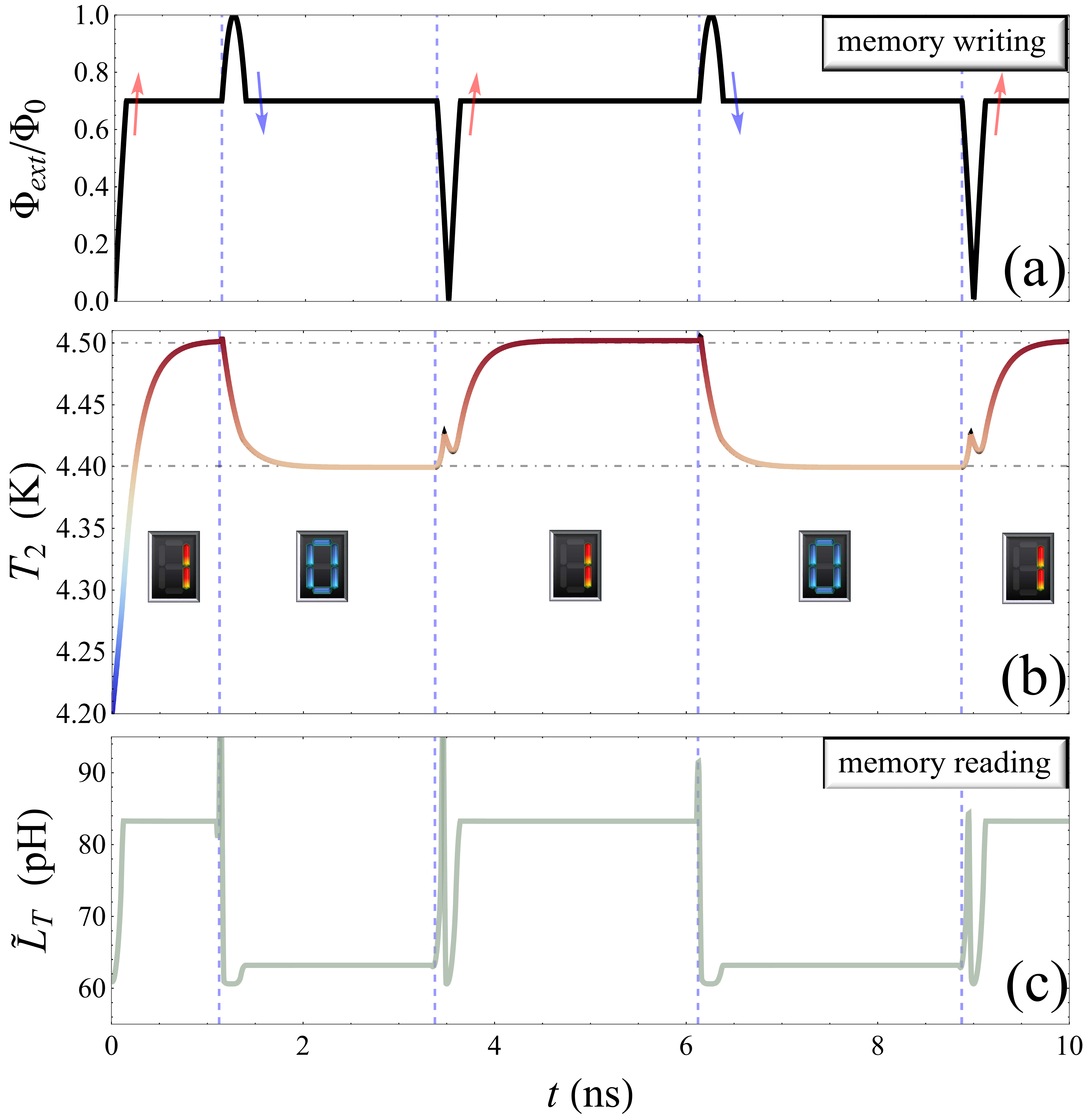}
\caption{(a) Driving flux giving an initial memory state 1 and four following switches, the latter indicated in all panels by blue dashed lines. A red (blue) arrow is used to indicate the increasing (decreasing) sweep direction of the drive when its value is frozen to the value $\Phi_{ref}=0.7\Phi_0$. (b) Time evolution of the temperature $T_2(t)$. Dot-dashed lines represent the asymptotic values approached by $T_2$ in the two cases. (c) Time evolution of the effective inductance, $\widetilde{L}_{T}$, see Eq.~\eqref{TankeffectiveInductance}, of the tank circuit coupled to the SQUID used for the memory state readout, as $L_T=100\;\text{pH}$.}
\label{Fig03}
\end{figure}

We discuss now how to drive the SQUID to control the thermal memory.
We assume a drive $\Phi_{ext}(t\leq t_{ref})=\Phi_0|\sin(2\pi\nu t)|$ and $\Phi_{ext}(t> t_{ref})=\Phi_{ref}$, with $t_{ref}$ being the time at which $\Phi_{ext}(t_{ref})=\Phi_{ref}$ (i.e., the driving flux is frozen to the value $\Phi_{ref}$ for $t\geq t_{ref}$). Time-dependent drives behaving in this manner are shown as solid lines in top panels of Figs.~\ref{Fig02}(a) and~\ref{Fig02}(b). Here, the dotted line represents the function $|\sin(2\pi\nu t)|$ and the red (blue) arrow indicates the increasing (decreasing) sweep direction of the drive when its value is fixed. Let us set, for instance, $\Phi_{ref}=0.7\Phi_0$. As is clearly shown in Figs.~\ref{Fig01} and~\ref{Fig02}, within each drive period, the condition $\Phi_{ext}=\Phi_{ref}=0.7\Phi_0$ occurs twice, once preceding and once following a phase swap. 

The temperature $T_2$ evolves differently according to the increasing, i.e., forward, or decreasing, i.e., backward, sweep direction of the driving flux when its value is kept fixed to $\Phi_{ref}$, as shown in the two driving protocols in Figs.~\ref{Fig02}(a) and~\ref{Fig02}(b), respectively. Although the driving flux finally assumes the same value, in the two cases $T_2$ tends to rapidly converge to different steady temperatures, $T_{2,\bit{1}}\simeq4.5\;\text{K}$ and $T_{2,\bit{0}}\simeq4.4\;\text{K}$ (indicated by two horizontal dot-dashed lines in Fig.~\ref{Fig02}). Hereafter, a subscript enclosed in a rectangle indicates the logical state associated to a specific value of an observable. 
Then, in an inductive SQUID the phase bistability, i.e., the $\pi$-swap of $\varphi$, reflects on a temperature bistability. Accordingly, we suggest a thermal memory in which the logic states 0 and 1 are defined by these two distinct values of the electronic temperature $T_2$. The writing operation of these states is performed through the driving flux.

The switch between the logical thermal memory states can be done over a short timescale, as shown in Fig~\ref{Fig03}, by driving the system through a $\pi$-swap and fixing it again as the $\Phi_{ref}$ value is reached anew. Accordingly, a memory state switch corresponds to a change in the sweep direction of the driving flux. Fig.~\ref{Fig03}(a) shows a magnetic drive giving an initial state 1 and four subsequent switches, the latter indicated by blue dashed lines. 
The temperature $T_2(t)$ increases from the value $T_2(0)=T_{bath}$, and the logic state 1 is rapidly reached, i.e., $T_{2}\simeq4.5\;\text{K}$, see Fig.~\ref{Fig03}(b). After each switch, the temperature $T_2(t)$ follows a transient regime and then, as the driving flux is kept fixed again, it exponentially approaches the steady temperature which is distinctive of the succeeding logic state, see Fig.~\ref{Fig03}(b). 

\emph{Memory switching time}. $-$ The time of the memory writing operation can be evaluated as the characteristic time of this exponential process, namely, $\tau_{wr}\sim0.2\;\text{ns}$. Markedly, a quite good estimate of this time results from first-order expanding the heat current terms in Eq.~\eqref{ThermalBalanceEq} (see Appendix~\ref{AppA}). In doing so, it derives from Eq.~\eqref{ThermalBalanceEq} that $C_v\frac{\mathrm{d} \Delta T_2}{\mathrm{d} t}=(K-G)\Delta T_2$ (see Appendix~\ref{AppA}), where $G$ and $K$ are the electron-phonon~\cite{Vir17} and electron~\cite{Mar14} thermal conductances of the JJs, respectively, and $\Delta T_2(t)$ represent the distance between $T_2(t)$ and its steady value, $T_{2_s}$. Therefore, $T_2$ exponentially approaches $T_{2_s}$, and the characteristic time of this process, i.e., the memory switching time, is $\tau_{sw} =C_v/(G-K)$. For $T_{2_s}=4.5\;\text{K}$, we obtain $\tau_{sw}\sim0.1\;\text{ns}$.

\begin{figure}[t!!]
\centering
\includegraphics[width=\columnwidth]{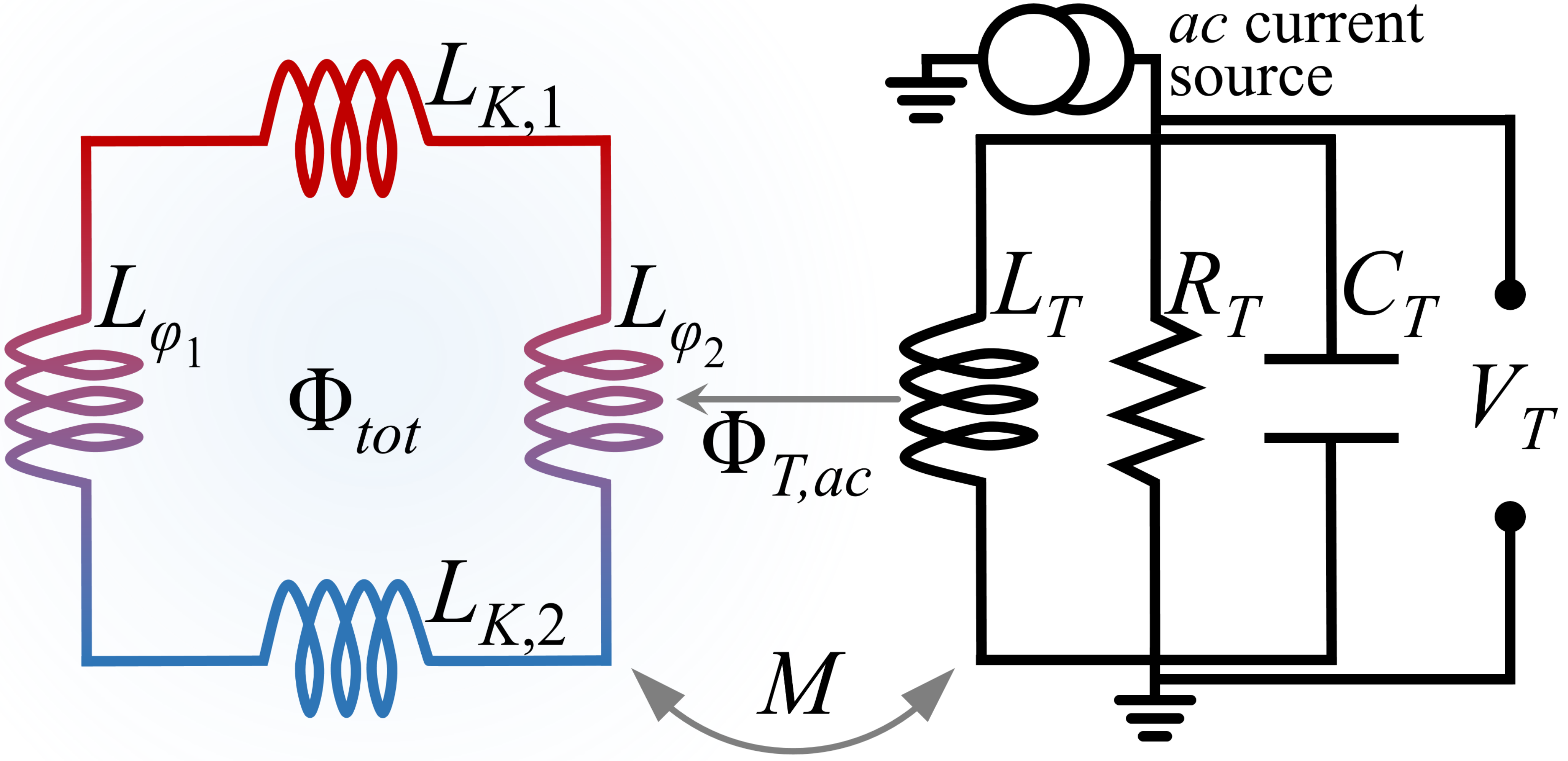}
\caption{Thermally biased SQUID (on the left) inductively coupled to a resonant tank circuit (on the right) characterized by an inductance, a resistance, and a capacitance $L_T$, $R_T$, and $C_T$, respectively. The total flux through the SQUID is $\Phi_{tot}=\Phi+\Phi_{T,ac}$ and $M$ is the mutual inductance of the system. }
\label{Fig07}
\end{figure}

\emph{The readout}. $-$ Recently, it has been shown that that real time temperature measurements in superconducting circuits at nanosecond scale are possible~\cite{Zgi17,Wan17}. 
Due to these technical achievements, the readout of the memory state can be done directly with time-dependent calorimetric measurements~\cite{Pek13,Gas15,Gia15,Sai16,Zgi17,Wan17}.
Hereafter, we do not specifically address those proposals and their applicability to the proposed thermal memory, but we suggest an alternative non-invasive indirect procedure. This readout scheme is based on the measurement of the effective inductance $\widetilde{L}_T$ of a tank circuit inductively coupled to our electrically-open double junction SQUID, as depicted in the equivalent electrical circuit schematically shown in Fig.~\ref{Fig07}. The state of the system can be measured via the effective resonance frequency of the tank circuit $\widetilde{f}_T=1\Big /\left ( 2\pi\sqrt{\widetilde{L}_T C_T} \right )$, where $\widetilde{L}_T$ is the effective inductance of the tank circuit (see Appendix~\ref{AppC}) and reads 
\begin{equation}
\widetilde{L}_T=L_T\left [1-\textsc{k}^2\frac{L}{L_J(\varphi_1,\varphi_2)+L} \right ],
\label{TankeffectiveInductance}
\end{equation}
where $\textsc{k}$ is the coupling coefficient defined by $M^2=\textsc{k}^2L_TL$, $M$ is the mutual inductance of the system, $C_T$ and $L_T$ are the capacitance and the inductance of the tank circuit, respectively. In Eq.~\eqref{TankeffectiveInductance}, $L$ is the ring inductance [see Eq.~\eqref{beta_T}], $L_J(\varphi_1,\varphi_2)=4L_{\varphi_1}L_{\varphi_2}/(L_{\varphi_1}+L_{\varphi_2})$ is the Josephson contribution, with 
\begin{equation}
L_{\varphi_i}(t)=\frac{\Phi_0}{2\pi}\frac{1}{I_{c,i } (T_1,T_2(t))\cos\left [\varphi_i (t) \right ]}
\label{JJKinInductance}
\end{equation}
being the Josephson inductance of the $i$-th JJ.
Assuming $C_T=1\;\text{nF}$, $L_T=100\;\text{pH}$, and $\textsc{k}=0.9$, we obtain for the logic state 1 the steady value $\widetilde{L}_{T,\bit{1}}\simeq83\;\text{pH}$ and a corresponding resonant frequency $\widetilde{f}_{T,\bit{1}}\simeq0.55\;\text{GHz}$, while for the state 0 we obtain $\widetilde{L}_{T,\bit{0}}\simeq63\;\text{pH}$ and $\widetilde{f}_{T,\bit{0}}\simeq0.63\;\text{GHz}$, see Fig.~\ref{Fig03}(c). A cavity with a modest quality factor, $Q\sim10$, should be able to resolve these two memory states, even if it is worthwhile to note that this $Q$ value needs to be increased considerably when the coupling term $\textsc{k}$ is reduced. 
In this scheme, the total flux $\Phi_{tot}$ through the SQUID is the sum of the flux $\Phi$, see Eq.\eqref{TotalFlux}, and the probing flux $\Phi_{T,ac}$ due to the tank circuit. For a proper operating point $\Phi_{ref}$ of the external flux, the oscillating component $\Phi_{T,ac}$ should be kept sufficiently small to avoid unwanted memory switches.
We stress that the main contribution to the difference in the effective inductance between the two thermal states is given by the Josephson terms, see Eq.~\eqref{JJKinInductance}, and increases with increasing the temperature difference between the memory states. 
Markedly, this detection scheme permits the tuning of the tank circuit effective resonant frequency by tuning the tank circuit parameters. 
Finally, the $L_{\varphi}$'s values can be increased, and therefore the visibility in the dispersive mode enhanced, by changing the critical current, namely, by increasing the JJ's resistance $R_j$. Nevertheless, the higher the $R_j$ values, the lower the temperature difference between the two logic states, so an optimal point has to be found. 

Notably, the thermal memory shows remarkable robustness against environmental disturbance (see Appendix~\ref{AppB}). In fact, our numerical calculations show that thermal Johnson–Nyquist noise currents~\cite{Bar82} only slightly affects the steady values of the temperatures, while the overall thermal dynamics remains essentially unchanged also in a stochastic Langevin approach.

\section{Conclusions}
\label{Conclusions}\vskip-0.2cm

In conclusion, we have discussed the bistable thermal behavior of a flux-controlled inductive SQUID and designed a superconducting fast thermal memory based on this effect. We propose a feasible fast readout scheme based on the reading of the effective resonance frequency of a tank circuit inductively coupled to the SQUID. The writing operation of the thermal memory states is performed through the driving flux. Interestingly, a writing time of the memory states on the order of $\sim0.2\;\text{ns}$ is in principle achievable. This is at least seven orders of magnitude faster than any other thermal memory proposed up to now. 

The proposed memory is well-placed in the context of superconducting memory elements~\cite{Mat80,Lik12,Rya12,Peo14,Sal17,GuaSol17,Mur17}, significantly pushes forward the thermal memory concept and has a strong relevance in the context of thermal devices.

In addition to the memory encoding, this device will use the heat produced by superconducting electronic circuits as a consequence of the computation to encode memory functionalities. Then, the proposed memory effectively paves the way for a new generation of fast thermal technology, including fundamental devices~\cite{ForGia17} and logic gates~\cite{Pao17}.

We acknowledge P. Virtanen, F. Paolucci, and G. Marchegiani for fruitful discussions.
C.G. and F.G. acknowledges the European Research Council under the European Union's Seventh Framework Program (FP7/2007-2013)/ERC Grant agreement No.~615187-COMANCHE for partial financial support. 
C.G. and P.S. have received funding from the European Union FP7/2007-2013 under REA Grant agreement No. 630925 -- COHEAT and from MIUR-FIRB2013 -- Project Coca (Grant No.~RBFR1379UX). 
A.B. acknowledges the Italian’s MIUR-FIRB 2012 via the HybridNanoDev project under Grant no. RBFR1236VV and CNR-CONICET cooperation programme ``Energy conversion in quantum nanoscale hybrid devices''.
M.D.V. acknowledges support from Department of Energy under Grant No.~DE-FG02-05ER46204.

\appendix

\section{Estimate of the characteristic switching time of the memory}
\label{AppA}

\begin{figure}[t!!]
\centering
\includegraphics[width=0.49\textwidth]{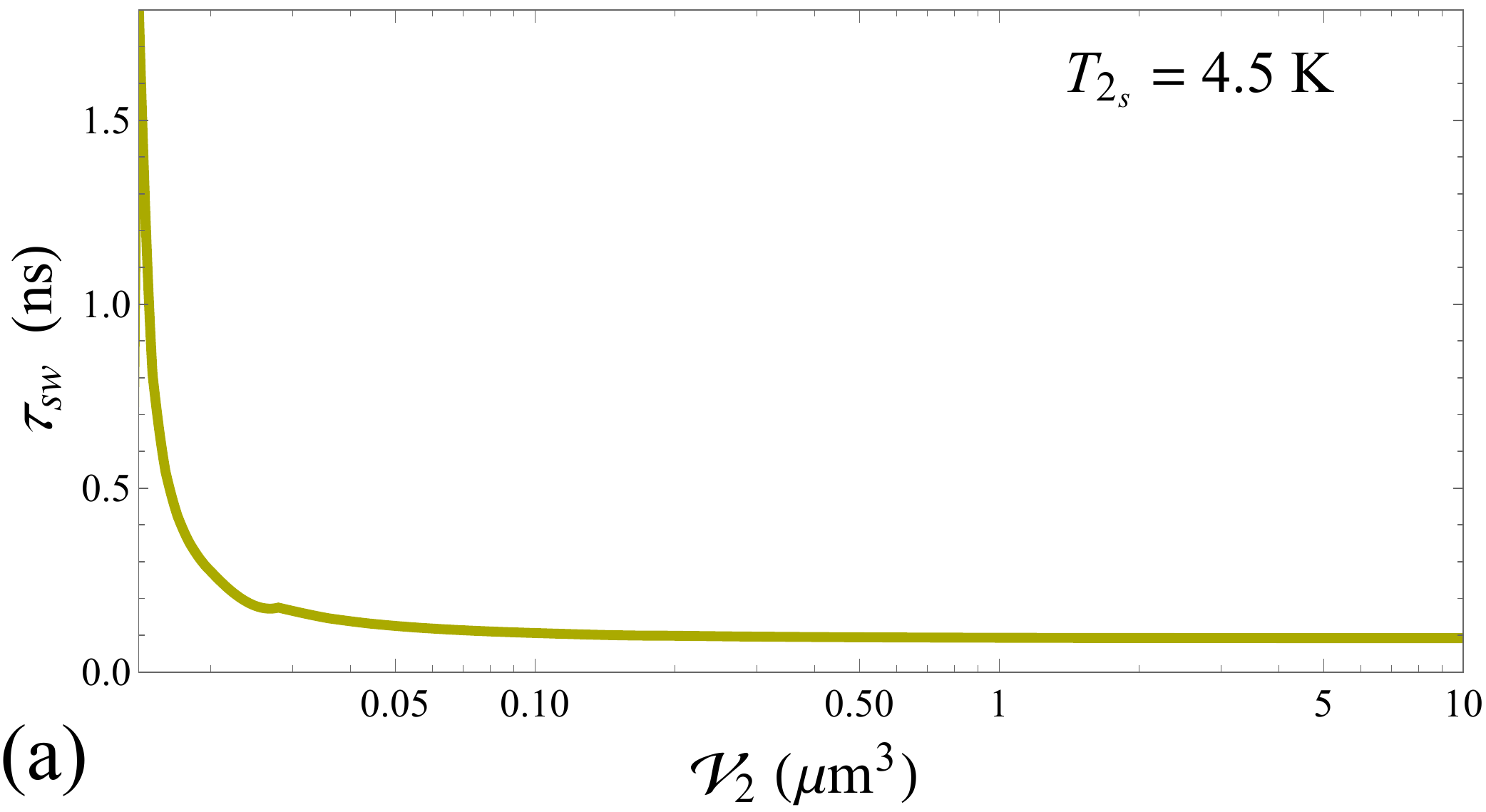}\\
\includegraphics[width=0.49\textwidth]{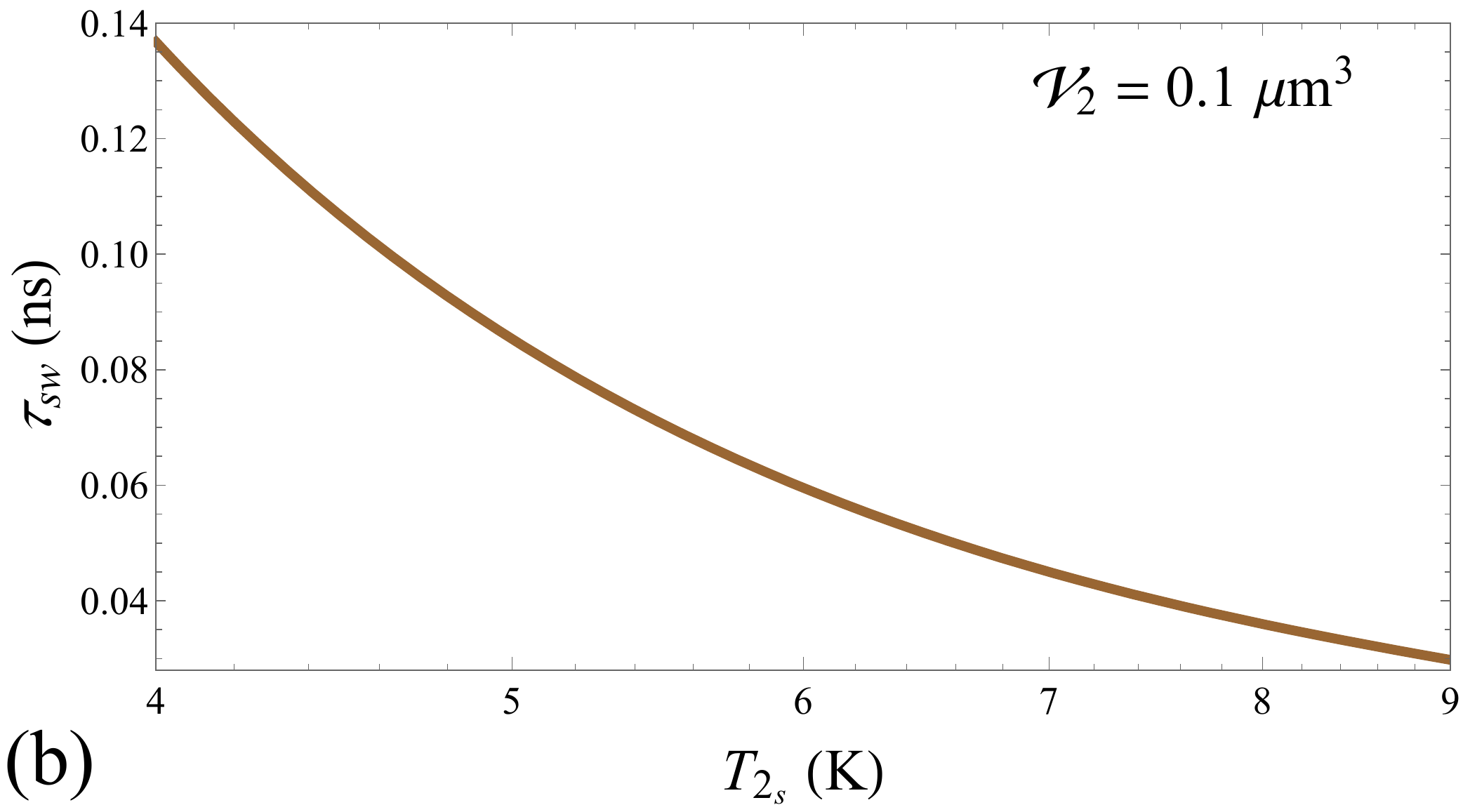}
\caption{Characteristic time, $\tau_{sw}$, as a function of both the volume $\mathcal{V}$, for $T_{2_s}=4.5\;\text{K}$ and $\Phi\simeq0.41\Phi_0$, (a), and the temperature $T_{2_s}$, for $\mathcal{V}=0.1\;\mu\text{m}^3$ and $\Phi=0.7\Phi_0$, (b). The values of the other parameters are the same used in the manuscript.}
\label{FigSM03}
\end{figure}
\begin{figure*}[ht!!]
\centering
\includegraphics[width=0.4\textwidth]{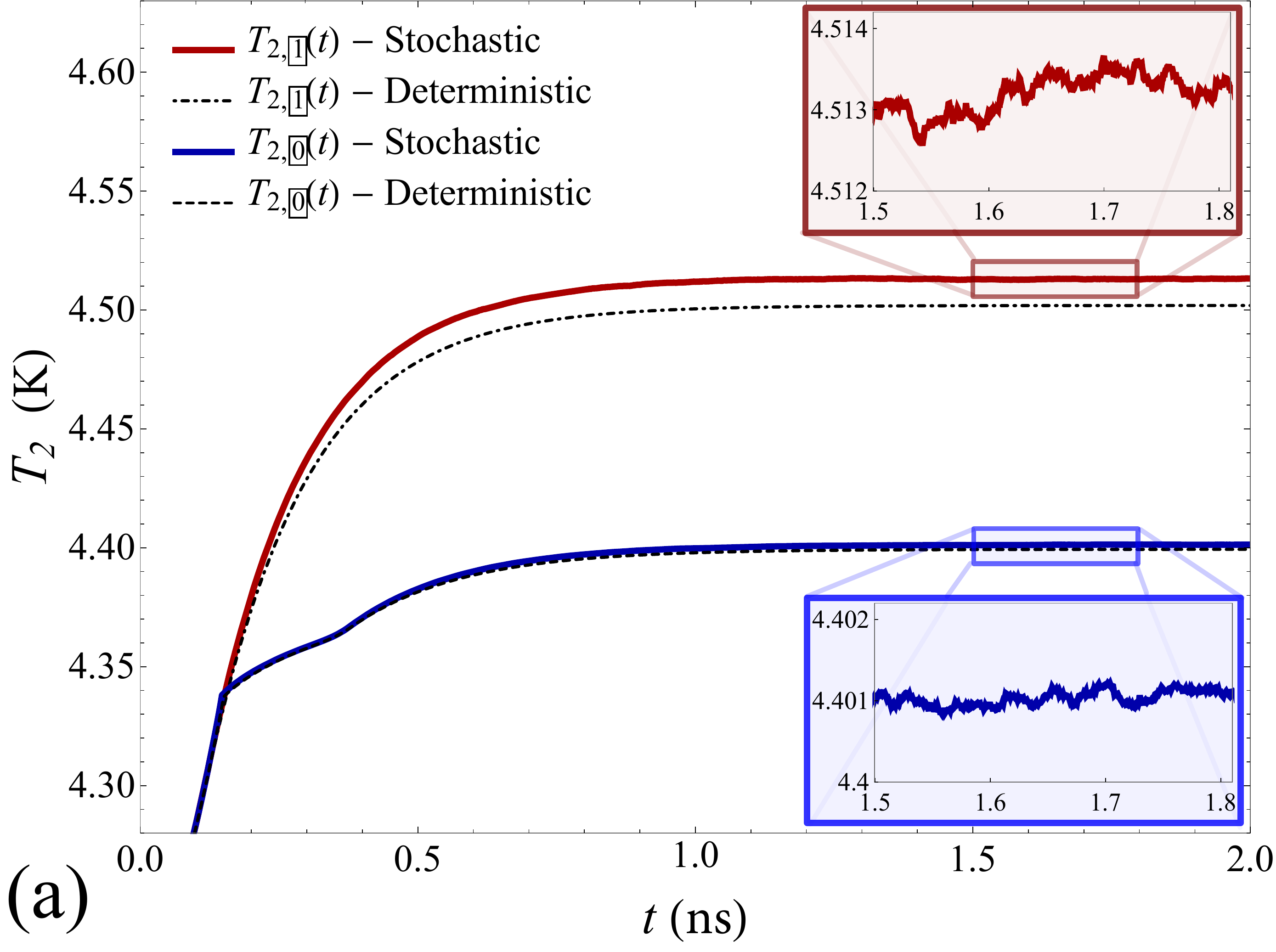}
\includegraphics[width=0.4\textwidth]{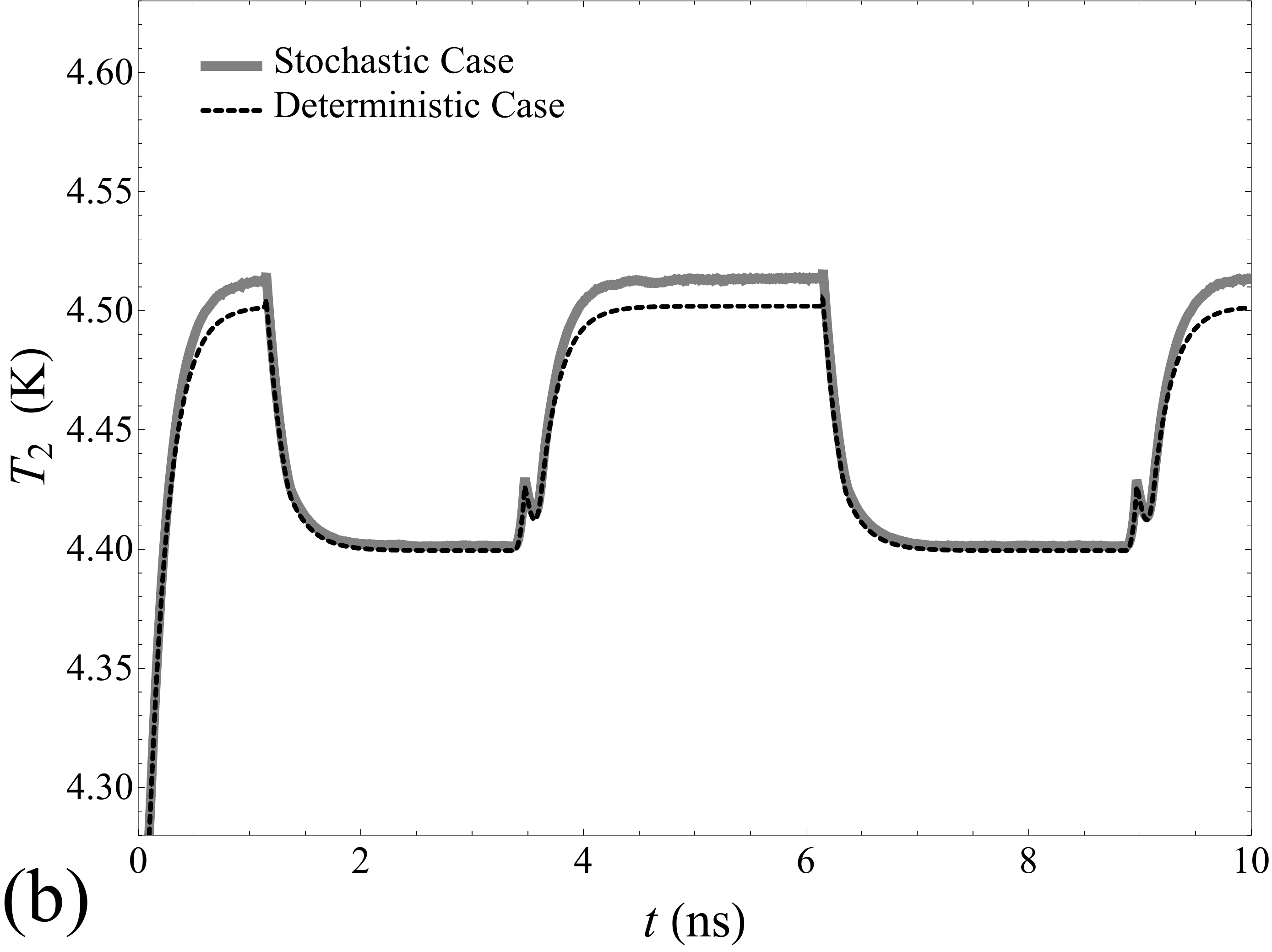}
\caption{(a) Time evolution of the temperature $T_2(t)$ as the thermal fluctuations are taken into account, for $\Phi_{ext}(t\geq t_{i})=\Phi_{ref}$, where $t_i$ is the time for the magnetic flux $\Phi_{ext}$ to reach the value $\Phi_{ref}=0.7\Phi_0$ during the forward and backward sweeps, for $\nu=1\;\text{GHz}$. The magnifications shown in insets allow to appreciate the stochastic fluctuations of $T_2$. (b) Evolution of the temperature $T_2(t)$ as the thermal fluctuations are taken into account, when several memory switches are driven by the external magnetic flux. In both panels, dashed curves represents the temperatures computed in the deterministic approach. The values of the other parameters are the same used in the manuscript.}
\label{FigSM04}
\end{figure*}

To give a qualitative estimate of the switching time, i.e., the time that the memory needs to change its state, we first-order expand the heat currents in the thermal balance equation Eq.~\eqref{ThermalBalanceEq} around the steady temperature $T_{2_s}$, obtaining the equation
\begin{eqnarray}\nonumber
&&P_t\left ( T_1,T_{2_s} \right )+\left .\frac{\partial P_T}{\partial T_2} \right |_{T_{2_s}}[T_2(t)-T_{2_s}]-P_{eph,2}\left ( T_{2_s},T_{bath}\right )\\
&&-\left .\frac{\partial P_{eph,2}}{\partial T_2} \right |_{T_{2_s}}[T_2(t)-T_{2_s}]=C_v(T_{2_s})\frac{\mathrm{d} T_2(t)}{\mathrm{d} t}.
\label{FirstOrderExpThermalBalance1}
\end{eqnarray}
According to the stationary thermal balance equation, i.e., $P_t\left ( T_1,T_{2_s} \right )-P_{eph,2}\left ( T_{2_s},T_{bath}\right )=0$, Eq.~\eqref{FirstOrderExpThermalBalance1} becomes
\begin{eqnarray}\nonumber
&&C_v(T_{2_s})\frac{\mathrm{d} T_2(t)}{\mathrm{d} t}\text{=}\left (\left .\frac{\partial P_t}{\partial T_2} \right |_{T_{2_s}}-\left .\frac{\partial P_{eph,2}}{\partial T_2} \right |_{T_{2_s}} \right )[T_2(t)-T_{2_s}]\\
&&=\left \{K(T_{2_s})-G(T_{2_s}) \right \}[T_2(t)-T_{2_s}],
\label{FirstOrderExpThermalBalance2}
\end{eqnarray}
where
\begin{eqnarray}
&&G(T)=\left .\frac{\partial P_{eph}}{\partial T_e} \right |_{T_e=T}=\\\nonumber
&&=\frac{5\Sigma\mathcal{V}}{960\zeta (5)k_B^6T^6}\iint_{-\infty}^{\infty}\frac{dEd\varepsilon E\left | \varepsilon \right |^3M^i_{E,E-\varepsilon }}{\sinh\frac{\varepsilon }{2k_BT}\cosh\frac{E }{2k_BT}\cosh\frac{E-\varepsilon }{2k_BT}}
\label{ephconductance}
\end{eqnarray}
is the electron-phonon thermal conductance~\cite{Vir17}, and 
\begin{eqnarray}\nonumber
&&K(T)=\left .\frac{\partial P_t}{\partial T_i} \right |_{T_i=T}=\\\nonumber
&&=\frac{1}{2e^2k_BT^2R_1}\int_{0}^{\infty}\frac{d\varepsilon \varepsilon^2}{\cosh^2\frac{\varepsilon }{2k_BT}}\Bigg [\mathcal{N}_1(\varepsilon,T)\mathcal{N}_2(\varepsilon,T)(1+a)-\\
&&-\mathcal{M}_1(\varepsilon,T)\mathcal{M}_2(\varepsilon,T)\sqrt{1+a^2+2a\cos\left ( \frac{2\pi \Phi}{\Phi_0} \right )}\Bigg ]
\label{econductance}
\end{eqnarray}

is the electron thermal conductance~\cite{Mar14} [here, $a=I_{c_{2}}/I_{c_{1}}=R_1/R_2=(1-r)/(1+r)$]. 

By defining $\Delta T_2=T_2(t)-T_{2_s}$, Eq.~\eqref{FirstOrderExpThermalBalance2} becomes
\begin{equation}
C_v(T_{2_s})\frac{\mathrm{d} \Delta T_2}{\mathrm{d} t}=\left \{K(T_{2_s})-G(T_{2_s}) \right \}\Delta T_2.
\label{}
\end{equation}
This equation can be recast in
\begin{equation}
\frac{\mathrm{d} \Delta T_2}{\mathrm{d} t}=-\frac{\Delta T_2}{\tau_{sw} },
\label{FirstOrderExpThermalBalance4}
\end{equation}
whose solution is
\begin{equation}
\Delta T_2(t)=\Delta T_{20}\;e^{-\frac{t}{\tau_{sw}}} ,
\label{FirstOrderExpThermalBalance5}
\end{equation}
where we have defined the characteristic time
\begin{equation}
\tau_{sw} =\frac{C_v(T_{2_s})}{G(T_{2_s})-K(T_{2_s})}.
\label{tau}
\end{equation}
For $T_{2_s}=4.5\;\text{K}$ and $\Phi_{ext}=0.7\Phi_0$, with $\Phi(t\gg1)\simeq0.41\Phi_0$ being the stationary value approached by the total flux, we obtain $\tau_{sw}\simeq0.1\;\text{ns}$, namely, a value roughly comparable with the time $\tau_{wr}$ discussed in the manuscript.

We note that a negative value of the switching time means that at the linear order the system is unstable. Anyway, the stability is recovered by considering the non-linear corrections. In order to optimize the switching times, we fixed the operating temperatures so that $G\gtrsim K$ and the system is stable at the linear order. 

Interestingly, the characteristic time $\tau_{sw}$ behaves counterintuitively by varying the volume $\mathcal{V}$ (in the following, we are assuming that any change in the volume $\mathcal{V}$ doesn't affect the JJ's surface area and, therefore, the JJs normal resistance). As is shown Fig.~\ref{FigSM03}(a), by increasing the volume, $\tau_{sw}$ approaches a steady value, while it tends to increase for small $\mathcal{V}$. To explain this trend, we observe that only $C_v$ and $G$ linearly depend on $\mathcal{V}$, so that, if we define $C_v=\mathcal{V}\widetilde{C_v}$ and $G=\mathcal{V}\widetilde{G}$, from Eq.~\eqref{tau} one obtains $\tau_{sw}=\widetilde{C_v}/(\widetilde{G}-K/\mathcal{V})$, according to which $\tau_{sw}\to\widetilde{C_v}/\widetilde{G}$ by increasing $\mathcal{V}$. A reduction of the island volume degrades the time performance of the memory element.

The behaviour of the characteristic time as a function of $T_{2_s}$, for $\mathcal{V}=0.1\;\mu\text{m}^3$ and $\Phi=0.7\Phi_0$, is shown in Fig.~\ref{FigSM03}(b). The characteristic time slightly reduces by increasing the temperature $T_{2_s}$, such that $\tau_{sw}\sim0.058\;\text{ns}$ for $T_{2_s}=6\;\text{K}$ and $\tau_{sw}\sim0.036\;\text{ns}$ for $T_{2_s}=8\;\text{K}$, since the electron-phonon relaxation process becomes more effective. Anyway, an increase of the temperature will result also in a reduction of the temperature difference between the two memory states increasing also the negative effects of the noise disturbance (see Appendix~\ref{AppB}).

\section{Effective coupling SQUID-tank circuit}
\label{AppC}

%
%
In order to implement the proposed readout scheme, we need to couple our SQUID with a tank circuit, as shown in Fig.\ref{Fig07}. In particular, as explained in the text the dispersive reading crucially depends on the effective inductance $\widetilde{L}_{T}$ of the tank circuit, which is modified by the different SQUID states. The method we discuss is inspired by the analysis presented in Ref.~\cite{Bar82} for a single-junction rf-SQUID. The difference here is that we have two junctions in the SQUID, whose phases are constrained by the flux quantization, see Eq.~\eqref{Fluxquantization}. Here we present how to derive the expression of the effective inductance $\widetilde{L}_{T}$, see Eq.\eqref{TankeffectiveInductance}. 

The tank circuit is coupled to the SQUID through the mutual inductance $M$, such that a fluctuation in the tank current, $I_T$, induces a fluctuation of the external flux, $\Phi_{ext}$, through the SQUID ring, namely, $\delta\Phi_{ext}=M\delta I_T$.
Consequently, a reactive circulating current is generated in the SQUID
\begin{equation}
\delta I_{circ}=\frac{\delta I_{circ}}{\delta \Phi_{ext}}\delta \Phi_{ext}=\frac{\delta I_{circ}}{\delta \Phi_{ext}}M\delta I_T,
\label{AppCEq03}
\end{equation}
which correspondingly induces a fluctuation of the total flux through the tank inductance, $\delta \Phi_T=L_T\delta I_T-M\delta I_{circ}$. Then, by using Eq.~\eqref{AppCEq03}, the effective inductance $\widetilde{L}_{T}$ reads
\begin{equation}
\widetilde{L}_{T}=\frac{\delta \Phi_T}{\delta I_T}=L_T\left ( 1- \frac{M^2}{L_T}\frac{\delta I_{circ}}{\delta \Phi_{ext}}\right ),
\label{AppCEq05}
\end{equation}
with $\delta I_{circ}/\delta \Phi_{ext}$ being the circulating current transfer function. The inverse of this transfer function can be obtained by differentiating Eq.~\eqref{TotalFlux}
\begin{equation}
\frac{\delta \Phi_{ext}}{\delta I_{circ}}=\frac{\delta \Phi}{\delta I_{circ}}+L.
\label{AppCEq07}
\end{equation}
The circulating current through the SQUID can be written as $I_{circ}=\left [I_{c_1}\sin\left ( \varphi+\phi \right )-I_{c_2}\sin\left ( \varphi-\phi \right )  \right ]/2$,
so that
\begin{equation}
\frac{\delta I_{circ}}{\delta \Phi}=\frac{\pi}{\Phi_0}\frac{\delta I_{circ}}{\delta \phi}=\frac{\pi}{2\Phi_0}\left [ I_{c_1}\cos\varphi_1+I_{c_2}\cos\varphi_2 \right ],
\label{AppCEq10}
\end{equation}
having used the relation $\delta\phi=\pi/\Phi_0\delta\Phi$. According to Eq.~\eqref{JJKinInductance}, the previous equation becomes
\begin{equation}
\frac{\delta I_{circ}}{\delta \Phi}=\frac{1}{4}\left ( \frac{1}{L_{\varphi_1}} +\frac{1}{L_{\varphi_2}} \right )=L_J(\varphi_1,\varphi_2)^{-1}.
\label{AppCEq11}
\end{equation}
Through Eqs.~\eqref{AppCEq07} and~\eqref{AppCEq11}, one obtains
\begin{eqnarray}
\widetilde{L}_{T}=L_T\left ( 1- \frac{M^2}{L_T}\frac{1}{L_J(\varphi_1,\varphi_2)+L}\right ).
\label{AppCEq13}
\end{eqnarray}
By assuming for the mutual inductance the expression $M=\textsc{k}\sqrt{L_TL}$, the effective tank inductance becomes
\begin{equation}
\widetilde{L}_{T}=L_T\left ( 1- \textsc{k}^2\frac{L}{L_J(\varphi_1,\varphi_2)+L}\right ),
\label{AppCEq15}
\end{equation}
where $\textsc{k}$ is the coupling coefficient. This equations resembles the result of Ref.~\cite{Bar82} for a single-junction SQUID coupled to a tank circuit, except that the Josephson contribution is represented by $L_J(\varphi_1,\varphi_2)$ defined in Eq.~\eqref{AppCEq11}.

\section{The noisy approach}
\label{AppB}

We consider two independent Johnson–Nyquist noise currents $I_{n_{1}}$ and $I_{n_{2}}$, with the usual white noise features~\cite{Bar82}
\begin{eqnarray}
\left \langle I_{n_{i}}(t) \right \rangle&=&0\\
\left \langle I_{n_{i}}(t)I_{n_{i}}({t}') \right \rangle&=&2\frac{k_BT}{R_i}\delta \left ( t-{t}' \right ),
\label{correlator}
\end{eqnarray}
in the RCSJ models of both junctions~\cite{Cla04,Gra16}
\begin{equation}
\begin{matrix}
\frac{I_{bias}}{2}+I_{circ}+I_{n_{1}}=\frac{\hbar }{2e}C_1\ddot{\varphi_{1}}+\frac{\hbar}{2e}\frac{\dot{\varphi_{1}}}{R_1}+I_{c_1}\sin\varphi_1\\ 
\frac{I_{bias}}{2}-I_{circ}+I_{n_{2}}=\frac{\hbar }{2e}C_2\ddot{\varphi_{2}}+\frac{\hbar}{2e}\frac{\dot{\varphi_{2}}}{R_2}+I_{c_2}\sin\varphi_2.
\end{matrix}
\end{equation}

With the aim of estimating the overall effect of the thermal fluctuations on the SQUID dynamics, we assume to first approximation that the temperature $T$ in Eq.~\eqref{correlator} is the higher temperature at play, namely, the temperature $T_1=6.5\;\text{K}$ of the hot electrode. 

Differences with respect to the deterministic case emerge by looking at the steady temperature $T_2$ in the memory states 0 and 1, i.e., $T_{2,\bit{0}}$ and $T_{2,\bit{1}}$, respectively. This is shown in Fig.~\ref{FigSM04}(a). We observe that the stochastic temperatures in the two cases converge to slightly higher values with respect to the temperatures computed in the deterministic case [indicated by dot-dashed and dashed black lines in Fig.~\ref{FigSM04}(a)].
The evolution of $T_2$ in the stochastic case, as several memory switches are guided by the magnetic drive, is shown in Figs.~\ref{FigSM04}(b). Despite the aforementioned shift, the overall dynamics remains unchanged. 

All these features can be explained with a detailed analysis of the stochastic regime, that we will investigate further in a following paper.


%

\end{document}